\documentclass{sig-alternate-10pt}

\usepackage{amssymb}
\usepackage{graphicx}
\usepackage{verbatim}
\usepackage[latin1]{inputenc}
\usepackage{paralist}
\usepackage{subfigure}
\usepackage{url}
\usepackage{times}
\def\sharedaffiliation{%
\end{tabular}
\begin{tabular}{c}}

\begin{document}

\title{Flow-level Characteristics of Spam and Ham}

\numberofauthors{3}
\author{
  \alignauthor Dominik Schatzmann\\
  \email{schatzmann@tik.ee.ethz.ch}
  \alignauthor Martin Burkhart\\
  \email{burkhart@tik.ee.ethz.ch}
  \alignauthor Thrasyvoulos Spyropoulos\\
  \email{spyropoulos@tik.ee.ethz.ch}
  \sharedaffiliation
  \affaddr{TIK-Report No. 291} \\
  \affaddr{Computer Engineering and Networks Laboratory}   \\
  \affaddr{ETH Zurich, Switzerland } \\
  \affaddr{29. August 2008}
}

\maketitle

\begin{abstract}

Despite a large amount of effort devoted in the past years trying to limit unsolicited mail, spam is still a major global concern. Content-analysis techniques and blacklists,  the most popular methods used to identify and block spam, are beginning to lose their edge in the battle. We argue here that one not only needs to look into the network-related characteristics of spam traffic, as has been recently suggested, but also to look deeper into the network core, in order to counter the increasing sophistication of spam-ing methods. Yet, at the same time, local knowledge available at a given server can often be irreplaceable in identifying specific spammers.

To this end, in this paper we show how the local intelligence of mail servers can be gathered and correlated \emph{passively} at the ISP-level providing valuable network-wide information. Specifically, we use first a large network flow trace from a medium size, national ISP, to demonstrate that the pre-filtering decisions of individual mail servers can be tracked and combined at the flow level. Then, we argue that such aggregated knowledge not only can allow ISPs to develop and evaluate powerful new methods for fighting spam, but also to monitor remotely what their ``own'' servers are doing.

\end{abstract}

\section{Introduction}

%
%

According to IronPort's 2008 Security Trend Report~\cite{ironport2008}, as much as 90\% of inbound mail is spam today. Morever, spam is no longer simply an irritant but becomes increasingly dangerous. 83\% of spam contains a URL. Thus, phishing sites and trojan infections of office and home systems alike are just one click away.

%
%
State-of-the-art spam filtering techniques are based on content analysis (e.g. SpamAssassin~\cite{SpamAssassin}), host reputation (SpamCop~\cite{SpamCop}, Spamhaus~\cite{Spamhaus}) or authentication services (SPF~\cite{wong2006spf}, DKIM~\cite{allman2006dkim}). Yet, with the increased use of attachment spam, success rate of content filtering decreases while its cost increases.
At the same time, despite the initial reported success of blacklists, it has been shown recently that their effectiveness is challenged by a new generation of low profile, short-lived, and highly dynamic botnets used to send spam~\cite{ramachandran2006unl}.

%
%

Consequently, new methods are needed to keep up with the spam arms race. One promising new approach to counteract this development is to focus on network-level characteristics of spammers. First, the network footprint of spamming activity can not be adapted as easily as the content of an email. Second, the rapid increase of spam traffic poses significant processing, storage, and scalability challenges for end-host systems, creating a need to at least perform some fast ``pre-filtering'' on the network level. Although the email content is not available in flow data, it can be used to profile potential spammers according to activities on other protocols and services offered. The network behavior of a legitimate email server is likely to be very different from a user PC acting as a botnet drone and sending spam.

%
%

To this end, a number of efforts have turned their attention towards network-level characteristics of spammers.  For example, in an interesting recent study Ramachandran et al.~\cite{ramachandran2006unl} have studied spam with respect to IP address and AS space and have demonstrated that clustering techniques can be applied to detect target domain patterns that are similar to those of spammers~\cite{ramachandran2007fsb}. Link analysis in MTA graphs has been applied by~\cite{desikan2004ant} to identify potentially malicious nodes. Gomes et al.~\cite{gomes2005cgt} showed that analysis of both static and evolving graph models allows to discriminate between socially unrelated spammer-receiver connections and socially related legitimate email traffic. Finally, an extensive characterization of spam traffic was given by Gomes et al.~\cite{gomes2004cas}. The authors highlighted differences between spam and legitimate emails with respect to a variety of aspects (not necessarily network-related), such as workload, temporal patterns, or sender and receiver distributions.

The majority of these works base their findings on mail server logs (e.g.~\cite{ramachandran2006unl, ramachandran2007fsb}) while a smaller minority on network flow data~\cite{clayton2006uer}. Log analysis provides detailed information about the local SMTP traffic \emph{and the local spam filtering knowledge} (i.e. which connections got accepted or rejected, along with the reasons thereupon) that goes far beyond simple reputation mechanisms, but is missing network-wide traffic statistics. Collecting server logs from many email servers can somewhat improve the situation. Yet, gathering logs from a large number of administration domains is a tedious process also involving privacy issues. On the other hand, flow data contains information for a much larger number of hosts, collected passively in the core of the network. Furthermore, flow traces are not limited to SMTP traffic and could allow one to correlate the behavior between different protocols or services. Yet, flow traces do not contain details about the content of SMTP sessions as they are restricted to packet header information. Most importantly, they do not right-away reveal the ``identity'' of a given connection with respect to the endpoints intentions (legitimate, spam, scanning, etc.).

We find that an ``either/or'' approach fails to combine the advantages of both methods, and suggest a different direction. Our proposal is to try to exploit and combine the valuable (and \emph{non-overlapping}) knowledge of local mail servers, on a network-wide scale, \emph{without actually collecting or looking at the individual server logs}. 

\begin{figure}[t]
	\centering
		\includegraphics[scale=0.30]{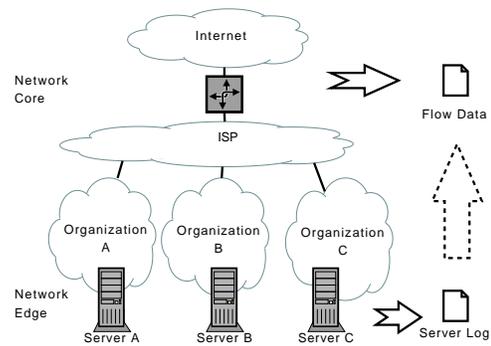}
	\caption{The ISP view of the network}
	\label{fig:ISPView}
\end{figure}

As a first step towards this goal, this paper presents two main contributions. \emph{First}, we show that email server's pre-filtering decisions can be accurately inferred directly from flow traces and therefore collected at the network core as illustrated in Fig.~\ref{fig:ISPView}. In particular, the distributions of SMTP flow properties, such as byte or packet count, reliably indicate which SMTP sessions were closed by the server before accepting contents. Based on this, MTA hosts can be classified as good or bad (sending spam) relying solely on packet header information. Our approach is validated using server logs and public black- and white-lists. In other words, \emph{spammer identities and the countermeasures deployed by mail servers become visible in flow traces}. 

To demonstrate the utility of these preliminary results, our \emph{second} contribution in this paper is to discuss a number of potential applications that could benefit from our approach, including anti-spam technique validation, global spam monitoring, server monitoring, and novel network-level spam countermeasures. In short, we believe that the rich information offered by flow traces will be crucial in the fight against current and next-generation spammers.

The rest of the paper is structured as follows. In the next section, we describe our methodology and datasets used in this study. Section~\ref{sec:results} we present our results with respect to the flow characteristics of accepted and rejected connections on single-server, and network-wide scales. Section~\ref{sec:discussion} describes a number of potential applications, as well as some limitations of our approach. Finally, we conclude this work in Section~\ref{sec:conclusion}.

\section{Methodology and Data}

\begin{figure}[t]
	\centering
		\includegraphics[scale=0.42]{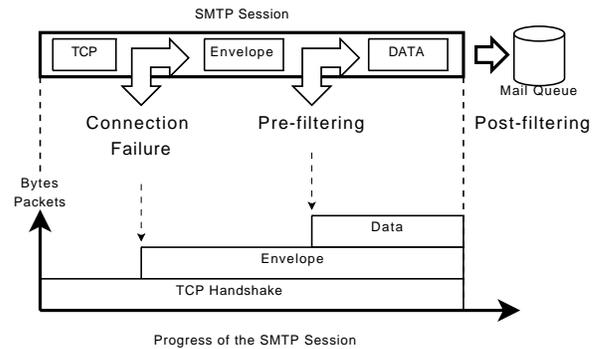}
	\caption{The three phases of email reception.}
	\label{fig:filteringPhases}
\end{figure}

In this section, we briefly describe our methodology and the datasets used for our study. As stated earlier, our goal is to combine the local knowledge available at individual email servers, on a network-wide level. To this end, we need first a model of the network-level behavior of an email server.

Consider the email reception process depicted in Fig.~\ref{fig:filteringPhases}. The first phase is the TCP three-way handshake initialized by the client. In the second phase, the email envelope is sent, according to the SMTP protocol~\cite{RFC2821}. At this stage, the server may apply certain pre-filtering rules to early close SMTP sessions from suspected spammers and in turn save valuable network, CPU and storage resources. If the server decides to go on, the email content is then transmitted in the third and last phase\footnote{Note that an SMTP session can be used to submit several messages consecutively, sending again an envelope and a sequence of data commands. Thus, a single SMTP session could be used to send both spam and ham messages. However, analyzing our university server log we found that only 1.5~\% of all SMTP connections contain more than one message. Therefore we assume that each SMTP session is used to send either spam or ham but not both.}.

Here, we are interested in the pre-filtering phase. During this phase, the server combines local and global knowledge to decide whether a new connection is a spam attempt. Local knowledge, for example, helps to decide whether the receiver address actually belongs to the server's domain (open relay attempt) or targets an existing user. Furthermore, greylisting could be used as a local reputation-based service. Global knowledge refers to the various open and commercial services, known as ``blacklists'', listing whether a sender has been known to send unsolicited content in the past. Only if the client does not ``clear'' the pre-filtering phase the SMTP session is closed by the server\footnote{We note that additional post-filtering is performed subsequently on accepted emails, based on their content. However, this post-filtering is not visible in the logs or flow data, and thus is orthogonal to our method. Our goal is to improve the pre-filtering phase, in order to alleviate the heavy load imposed on post-filtering.}.

We analyzed the logs of a university email server and show the top reject reasons in Table~\ref{tab:reject}. As much as 78.3\% of the SMTP sessions were rejected in the pre-filtering phase. Rejects based on local information (including many reasons subsumed in \textit{Others}) make up for roughly 45\% of the reject reasons. \emph{These results nicely illustrate the importance of local mail server knowledge for spam detection.}

\begin{table}[t]
	\centering
	\begin{tabular}{|l|r|}
	\hline
		Reject Reason & Percentage \\ 
	\hline
		Blacklist Spamhaus						& 37.5\% \\
		R-Addr: user unknown					& 15.6\% \\
		R-Addr: invalid domain				    & 18.3\% \\
		Helo: no FQDN							& 17.4\% \\
		Relay: denied							&  0.2\% \\
		\textit{Others}							& 11.0\% \\
	\hline
	\end{tabular}
	\caption{Reject reasons in pre-filtering.}
	\label{tab:reject}
\end{table}

Based on these findings, we are interested in answering the following questions: 
\begin{enumerate}

\item Do the various connections in the above log exhibit different enough flow characteristics to reliably identify the decision of the server (accept, reject) and the nature of a connection attempt (scan, spam, legitimate) without looking at the log?
     
\item Do these flow characteristics carry over into the network core when traffic from a large set of different email servers (with possibly different filtering policies or role) is aggregated?
    
\end{enumerate}

To answer both questions, we utilize NetFlow data collected by the border routers of a major national ISP serving more than 30 universities and government institutions. The IP address range contains about 2.2 million internal IP addresses and the traffic volume varies between 60 and 140 million NetFlow records per hour. We have used this data to try to isolate patterns indicating the mail servers decisions for SMTP flows.

To answer the first question (Section~\ref{sec:singleServer}), we have identified in our NetFlow trace the flows corresponding to the server log analyzed above. Preliminary results show that it is indeed possible to tell spammers apart from legitimate email senders by looking at feature distributions such as flow size, packet count and bytes per packet. This is feasible because in each phase of the email reception process, the number of transferred bytes and packets is significantly higher than during the previous phase. 

To answer the second question (Section~\ref{sec:networkwide}) we extend the scope to the entire network. In our flow traces, we see traffic from more than 50 mail servers. In this case, it would be impossible to get the logs of all servers and perform a similar study, due to administrative and privacy concerns. Instead, in order to validate our results in this second experiment, we classified flows according to the membership of the sender in either a black- or whitelist~\cite{Spamhaus, dnswl} and compared their statistics.

\section{Results}
\label{sec:results}

In this section, we present and discuss our preliminary results with respect to the flow characteristics of accepted and reject SMTP flows.

\begin{figure*}[htbp]
  \centering
  \subfigure[Flow size distribution]{
    \includegraphics[angle=270, scale=0.22]{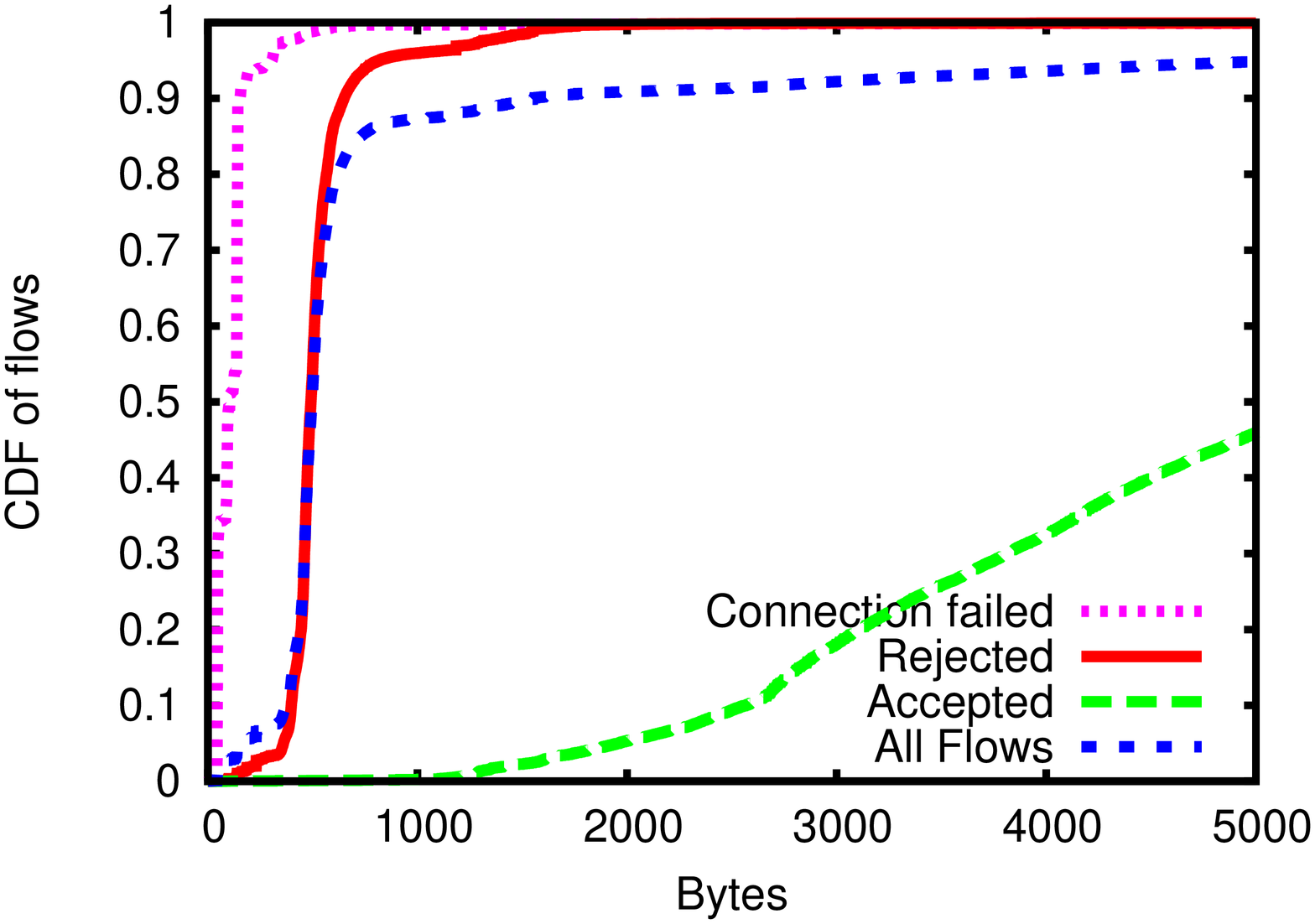}
    \label{fig:sizesLog}
  }
  \subfigure[Packet count distribution]{
    \includegraphics[angle=270, scale=0.22]{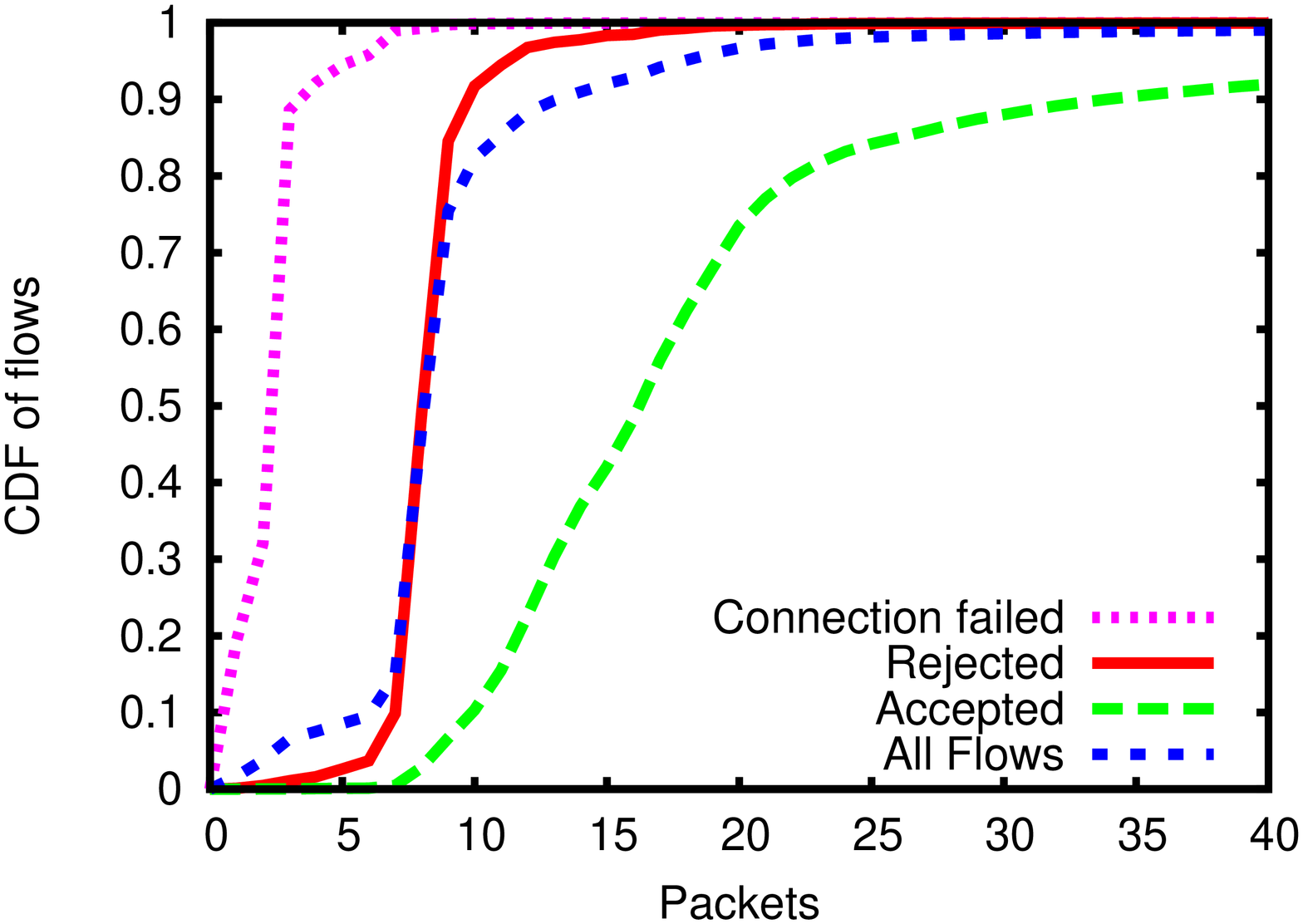}
    \label{fig:packetsLog}
  }
  \subfigure[Bytes per packet distribution]{
    \includegraphics[angle=270, scale=0.22]{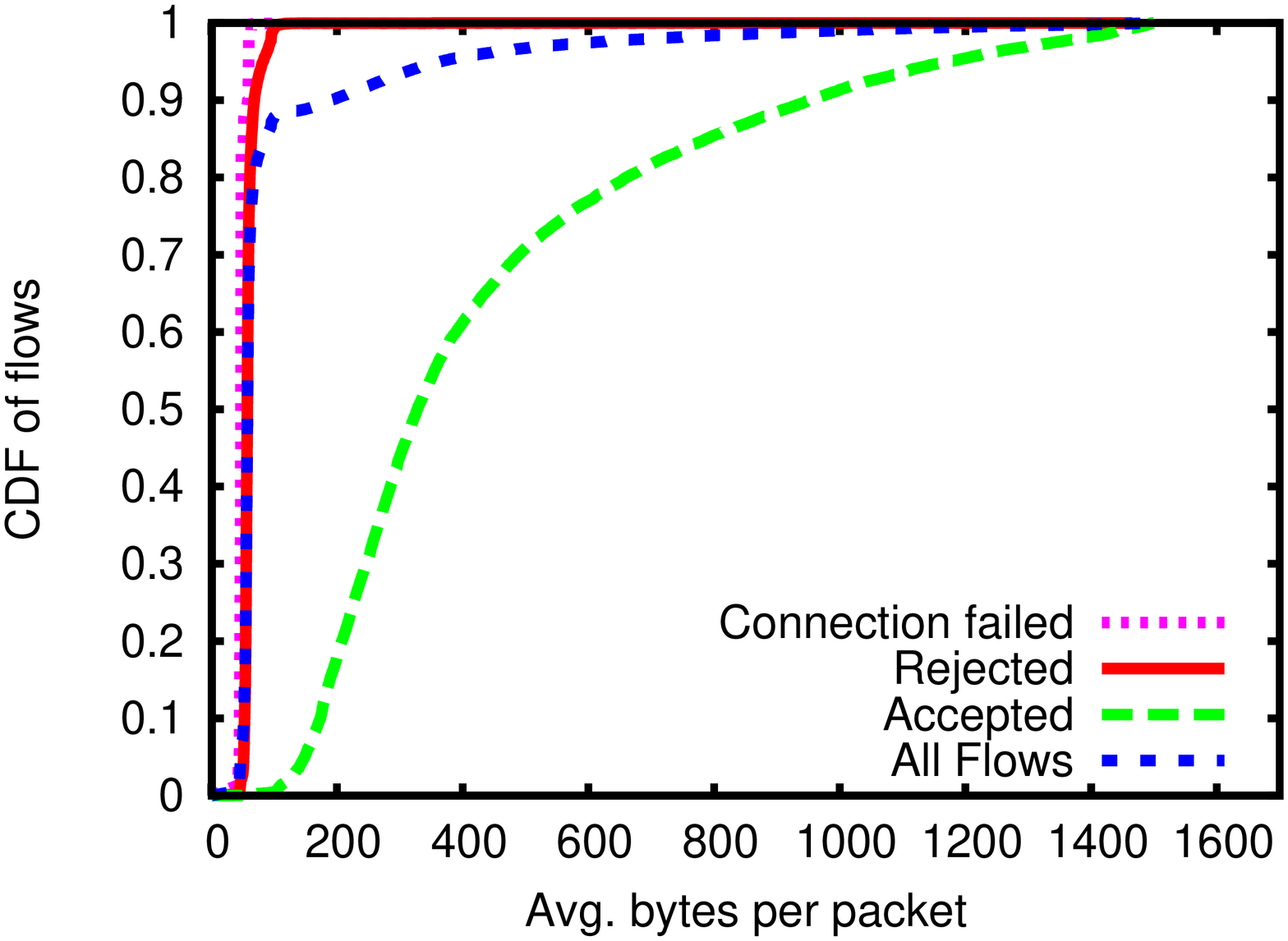}
    \label{fig:sizesLogBPP}
  }
  \caption{Feature distributions of flows to a single server.}
  \label{fig:singleServer}
\end{figure*}

\subsection{Single server characteristics}
\label{sec:singleServer}

We use the server logs to divide all flows into three classes, namely \emph{accepted}, \emph{rejected} and \emph{failed connection attempts}. If we can find a matching session for a given flow in the server log, we look into the respective log entry to decide whether is was rejected in pre-filtering or accepted. If no entry is present, the flow is labeled as a failed connection attempt.

Fig~\ref{fig:singleServer} shows the distributions of flows sizes (a), packet count (b), and bytes/packet (c) for flows arriving at the the mail server. It is obvious that each flow class has very specific characteristics. Consider, for example, the flow size seen in Fig.~\ref{fig:sizesLog}:

\begin{description}
\item[Failed connections:] Mainly very small flows as only a small number of packets is sent. 95\% of these flows have less than 300 bytes.
\item[Rejected:] Only 3\% of flows are below 300 bytes. Most flows are between 400 and 800 bytes. 99\% of flows have less than 1500 bytes.
\item[Accepted:] Only 5\% of flows have less than 1500 bytes. This distribution is dominated by email size distribution and reaches 50\% at around 5000 bytes. This is consistent with the findings of Gomes et al.~\cite{gomes2004cas}.
\end{description}

The CDF for ``all flows'' is a superposition of the three classes weighted by their relative probability of occurrence. All three classes are visible in the CDF even though it is dominated by the CDF of the rejected flows due to the fact that around 80\% of the flows are rejected.

Figure~\ref{fig:packetsLog} shows the same graph for packet count instead of byte count (flow size). The difference between rejected and failed connections is very strong. However, the packet count of accepted flows is not as distinctive as the flow size in Fig~\ref{fig:sizesLog}, i.e., the accepted curve is closer to the rejected curve.

The plots for bytes/packet in Fig~\ref{fig:sizesLogBPP} are similar to bytes per flow but are upper bounded by 1500 bytes (the typical MTU packet size of Ethernet). The average bytes per packet of accepted flows are dominated by the transportation of the email message whereas the rejected flows are dominated by the SMTP envelope traffic. The later is again dominated by sending only a small amount of data per packet, e.g. an email address. This result suggests that it is possible to detect the three different classes even if we have only one or two packets per flow, as it might be the case when packet sampling is applied.

Characteristic ranges from Fig.~\ref{fig:singleServer} are summarized in Table~\ref{tab:ranges}. These results demonstrate that it is possible to distinguish spam from ham, in the given mail server, with high confidentiality by looking at simple flow properties.

\begin{table}
	\centering
		\begin{tabular}{|l|ccc|}
		  \hline
			Category						& Bytes       & Packets  & Bytes/Packet \\
			\hline
		  failed conn.        & $<300$      & $<5$      & $<100$  \\
		  rejected				    & $300-1500$  & $5-10$    & $<100$  \\
		  accepted						& $>1500$     & $>10$     & $>100$  \\
		  \hline
		\end{tabular}
		\caption{Characteristic flow properties.}
		\label{tab:ranges}
\end{table}

\subsection{Network-wide characteristics}
\label{sec:networkwide}

The above classification was derived from the logs of a single server. The question arises whether the results can be extended to other servers in the network.
To address this question we now look at the total SMTP traffic of the 50 most active email servers of the network.

Because we could not get the server logs of all the servers we classify flows according to the membership of the sender in either a black- or whitelist. We are interested whether flows originating from blacklisted hosts exhibit characteristics similar to the rejected flows of the previous section. Similarly, whitelisted hosts should mainly send accepted traffic and thus exhibit according flow characteristics.

\begin{figure*}[htbp]
  \centering
  \subfigure[Network-wide flow sizes]{
    \includegraphics[angle=270, scale=0.22]{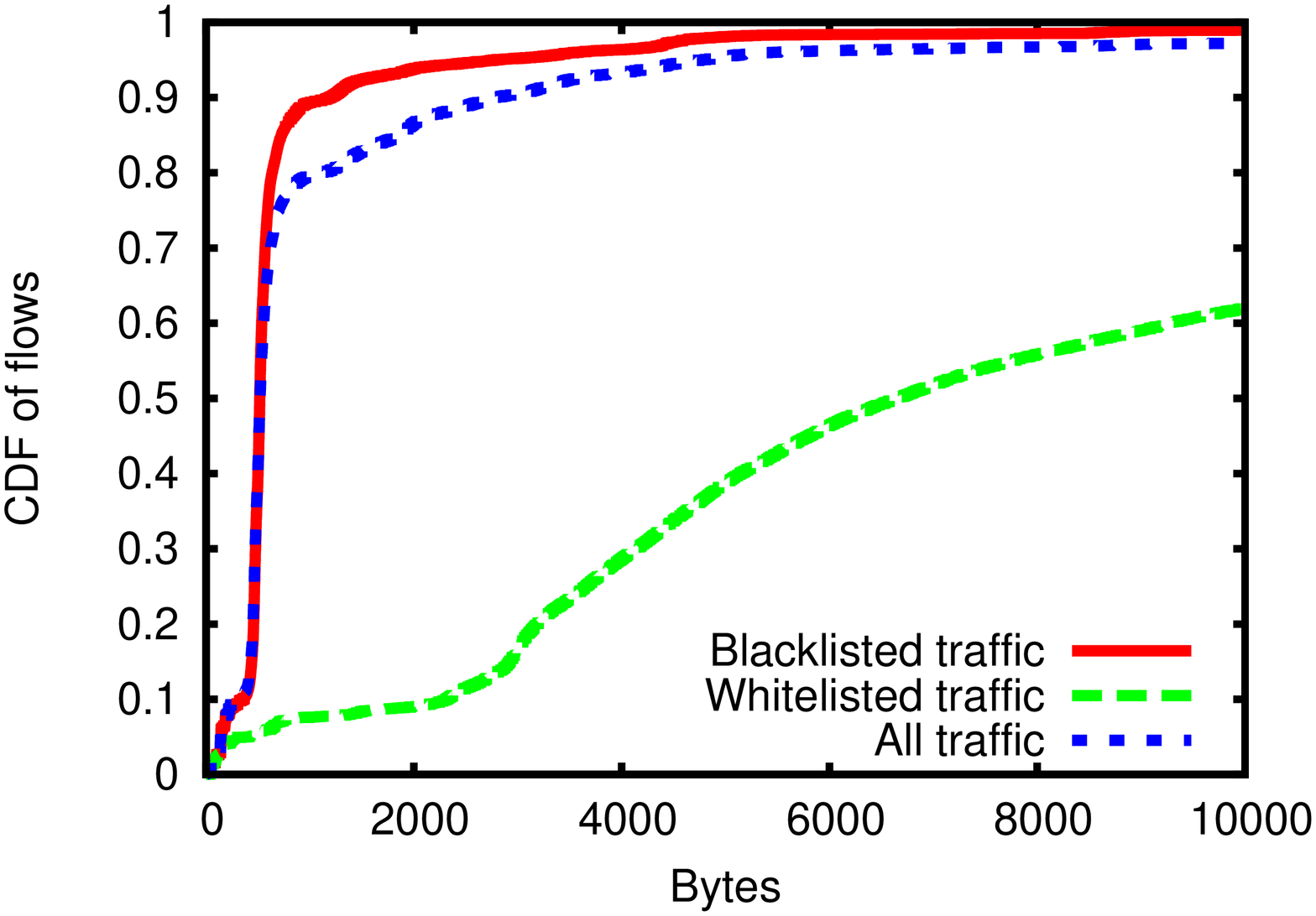}
    \label{fig:sizesList}
  }
  \subfigure[Flow sizes of different email servers]{
    \includegraphics[angle=270, scale=0.22]{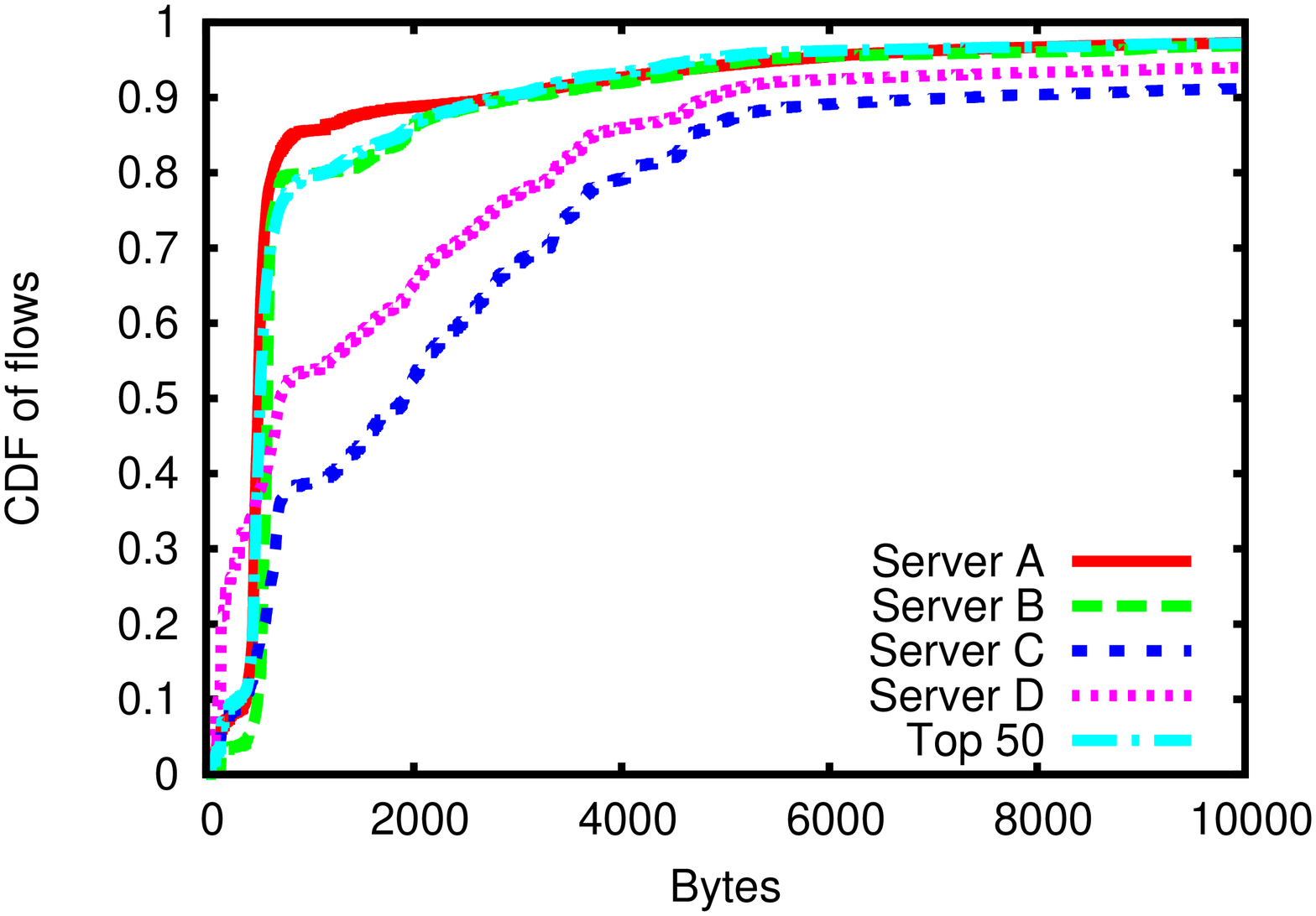}
    \label{fig:perserver}
  }
  \subfigure[SMTP traffic of the top server]{
    \includegraphics[angle=270, scale=0.22]{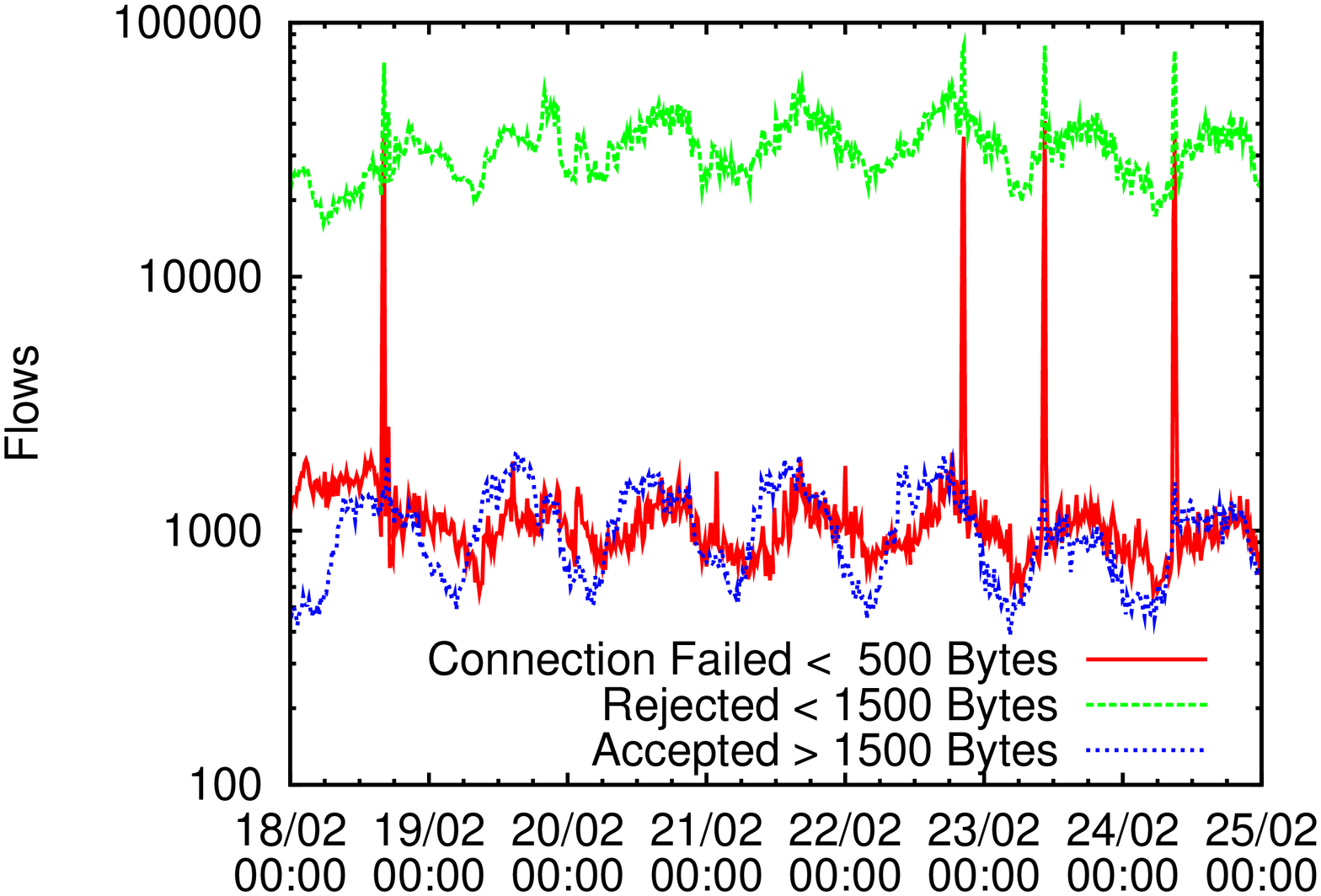}
    \label{fig:overtime}
  }
  \caption{Flow sizes in the entire network (a), for selected servers (b) and traffic overview (c). }
\end{figure*}

Exemplified for flow sizes, Figure~\ref{fig:sizesList} shows that, indeed, the shape of rejected and accepted flows is nicely reflected by the black-/whitelisted curves. The CDF of blacklisted traffic rises in the region of 300 bytes and achieves a ratio of 92\% at 1500 bytes. In contrast, the CDF ot the whitelisted traffic stays on a low level of 10\% up to 1500 bytes and starts to increase above 1500 bytes.

Yet, the fraction of whitelisted flows below 1500 bytes is not so small as for accepted traffic in Fig.~\ref{fig:sizesLog}. We assume this is due to applied greylisting in servers, which leads to a number of rejected flows even for whitelisted hosts. In addition, whereas the fraction of rejected flows smaller than 1500 bytes was 99\% in Fig.~\ref{fig:sizesLog}, we now have a slightly smaller ratio of about 92\% for blacklisted flows. This means that the ensemble of the top 50 servers has a less aggressive pre-filtering configuration than the single email server studied before. Also, if other servers use a blacklist that performs worse than Spamhaus, the blacklist used by our server, this effect would be expected. That also implies that less knowledgeable servers could benefit from the knowledge of servers that are better configured, that is, servers that exhibit a sharper reject curve. Again, the curve of all traffic is dominated by blacklisted flows.

Summing up, these results imply that the discriminating power of the three metrics (bytes per flow, packets per flow, and bytes per packet) with regard to spamming activity is also valid in a network-wide setting.

\section{Discussion and Applications}
\label{sec:discussion}

\paragraph{Impact}


In this paper, we established a passive measurement method for detecting the mail server's acceptance of email sending attempts. The method requires no additional communication or configuration changes on the mail server. To our best knowledge, this is the first method that scales to \emph{large networks} and \emph{passively} transfers knowledge about spam from the edge of the network (servers) to the core (as shown in Fig.~\ref{fig:ISPView}). Although the proposed method is quite simple and based on few basic flow characteristics, we believe it paves the way towards more sophisticated algorithms for network-level spam detection that could successfully complement existing methods.


Servers at the edge of the network have local knowledge, such as the user database or the email envelope, that is not accessible at the core of the network. Further on, servers correlate this knowledge with services (DBL, SPF) to improve the accuracy of their decision. With the presented method it is possible to concentrate the local decisions of several email servers in the core of the network. This approach can be interpreted as a distributed computer cluster where the email servers perform a detection algorithm on part of the global SMTP traffic and we are then collecting the results.

Mail servers only see their own traffic. In the core of the network we have the ``bird's eye view'' over the activity of all nodes. This allows to correlate the gained information from one email server with the information of many others. Further on, our approach is not restricted to SMTP traffic and, therefore, one can try to find suspicious behavior in other protocols, too (a subject of future work).

Although being only a first step, we believe the proposed method has important potential applications in production networks and also paves the way for future research in the area of network-level and network-wide characteristics of spam traffic. In the rest, we try to illustrate some of these applications.

\paragraph{ISP Use Case}

\emph{Network-wide spam monitoring:} By looking at the flow characteristics of incoming email traffic, an ISP is now able to \emph{estimate the total spam traffic} collected by its network. Furthermore, it is possible to focus on the statistics of a specific server and to \emph{detect potential performance problems}. For example, Figure~\ref{fig:overtime} shows the three categories of flows that arrive at an email server inside the network in 15 min. intervals. 
On 18. Feb at 6:00, the server rejects up to 80,000 flows. We think that this event corresponds to a spam campaign arriving at the server. During this time, the server is probably overloaded and starts rejecting connection attempts. Three similar events are visible at the end of the week. These events are visible on other severs as well and are therefore of a distributed nature. We believe that an ISP could monitor its network thus, and re-act in near real-time to such suspicious events.

As a second observation, note how the diurnal pattern of rejected/failed flows has an offset of approximately 6h compared to accepted flows. Since the local time zone of the server is GMT+1h, this indicates that most spam to that server is sent from time zone GMT-5h (including eastern USA).

\emph{Internal server monitoring:} In addition, the ISP could \emph{rate servers} by calculating the ratio of accepted vs. rejected flows for each server. This ratio could be used to to detect open relay servers or servers with a suboptimal pre-filtering process, triggered by an unusually high number of accepted flows. These misconfigured servers produce unnecessary and costly traffic. The ISP can now take counteractive measures. For example, Fig.~\ref{fig:perserver} shows the flow size curves for different mail servers in our network. Some of the servers perform a very aggressive pre-filtering (server A) while others do not seem to care too much about it (server C).

By monitoring the three categories of outgoing SMTP traffic, the ISP could infer the reputation of internal sources. For example, the reputation could be calculated as the ratio of accepted versus total flows. Based on this value, the ISP would be able to \emph{detected blacklisted internal servers} or servers abused for a spam campaign. With the same method, \emph{spamming botnet members} inside the network could possibly be detected by the ISP.

\emph{Rating external email sources:} Further the internal servers could be used as agents of a collaborative filtering system. The external mail senders could be rated based on the three categories of their incoming and outgoing SMTP traffic. This system could be used to possibly even filter out spam on the edge of the network. The rating of the external email sources could be published as a service for other email servers.

Given that an operator is already collecting flow data, e.g. for accounting, the data required by the outlined applications is already available. There is \emph{no need to correlate the traffic again} with different blacklists or other sources. The email servers have already done this work. To implement a similar service without flow data, the ISP would have to contact each mail server administrator to get access to their local server log. This process is bound to be cumbersome or even impossible due to privacy issues. With our approach only one sensor must be maintained.

\paragraph{Limitations}

Tracing the footprints of pre-filtering methods in network traffic has the inherent limitation that it only detects patterns if pre-filtering is applied by servers in the first place. However, in production environments it is the email administrator's intention to block spam as early as possible to save network, CPU and storage resources. Pre-filtering is the preferred choice to achieve this. Moreover, pre-filtering allows to reduce the amount of non-delivery messages to be sent~\cite{FDMailDDOS}. Therefore, we believe it is reasonable to rely on the presence of pre-filtering in production networks.

In principle, spammers could adapt to our method by prolonging the envelope phase, e.g. by sending multiple RCPT or HELO commands or providing very long sender email addresses. However, any deviation from normal SMTP behavior could again be detected and mail servers could enforce the shortness of the envelope as there is no need to be overly long. 

\section{Conclusion}
\label{sec:conclusion}

Mail server administrators are engaged in an arms race against spammers. They urgently need new approaches to fight state-of-the-art attachment spam increasingly originating from low-profile botnet spammers.

In this paper, we demonstrated that simple flow metrics, such as byte count, packet count, and bytes per packet, successfully discriminate between spam and ham flows when pre-filtering is deployed in mail servers. Thus, one could infer individual mail server's decisions with respect to the legitimacy and acceptance of a given connection. This allows an operator i) to concentrate dispersed mail server knowledge at the network core and ii) to passively accumulate network-wide spam statistics, profile filtering performance of servers, and rate clients. Thus, the advantages of flow and server log analysis finally meet at the network core. 
We believe this is an important step towards successfully fighting spammers at the network-level.

\bibliographystyle{abbrv}

\balancecolumns

\end{document}